\documentstyle[prd,aps,epsfig]{revtex}
\begin{document}
\def\lesssim{\mathrel{\hbox{\rlap{\hbox{\lower4pt\hbox{$\sim$}}}\hbox{$<$}}}}
\def\gtrsim{\mathrel{\hbox{\rlap{\hbox{\lower4pt\hbox{$\sim$}}}\hbox{$>$}}}}
\newcommand{\figdir}{./}
\setlength{\textwidth}{7.00in}
\setlength{\textheight}{9.0in}
\setlength{\evensidemargin}{-0.2in}
\setlength{\oddsidemargin}{-0.2in}
\setlength{\topmargin}{0.0in}
\input epsf
\draft
\renewcommand{\topfraction}{0.8}
\twocolumn[\hsize\textwidth\columnwidth\hsize\csname
@twocolumnfalse\endcsname \preprint{CITA-2000-16, SU-ITP-00-28
hep-ph/0011160, November, 2000}
\title { \Large  \bf The Development of Equilibrium After Preheating}
\author{Gary Felder$^{1,2}$ and Lev Kofman$^2$}
\address{${}^1$Department of Physics, Stanford University, Stanford, CA
94305, USA}
\address{${}^2$CITA, University of Toronto, 60 St George Str,
Toronto, ON M5S 3H8, Canada}
\date{November, 2000}
\vspace{-.1in}
\maketitle
\begin{abstract}
We present a fully nonlinear study of the development of
equilibrium after preheating. Preheating is the exponentially
rapid transfer of energy from the nearly homogeneous inflaton
field to fluctuations of other fields and/or the inflaton itself.
This rapid transfer leaves these fields in a highly nonthermal
state with energy concentrated in infrared modes. We have
performed lattice simulations of the evolution of interacting
scalar fields during and after preheating for a variety of
inflationary models. We have formulated a set of generic rules
that govern the thermalization process in all of these models.
Notably, we see that once one of the  fields is amplified through
parametric resonance or other mechanisms it rapidly excites other
coupled fields to exponentially large occupation numbers. These
fields quickly acquire nearly thermal spectra in the infrared,
which gradually propagates into higher momenta. Prior to the
formation of total equilibrium, the excited fields group into
subsets with almost identical characteristics (e.g. group
effective temperature). The way fields form into these groups and
the properties of the groups depend on the couplings between them.
We also studied the onset of chaos after preheating by calculating
the Lyapunov exponent of the scalar fields.
\end{abstract}

\pacs{
%PACS: 98.80.Cq
 \hskip 1.5cm  CITA-2000-61 \hskip 1.5cm SU-ITP-00-28 \hskip 3.0cm \
hep-ph/0011160}

\vskip2pc]

\section{Introduction}

The theory of inflation has been highly successful in explaining
many of the initial conditions for the
 Hot Big Bang model as well as providing
a mechanism by which the seeds of large scale structure were
formed. Typical models of inflation are based on the slow-roll
evolution of the homogeneous inflaton scalar field(s) $\phi$.
Inflation ends when the slow-roll regime is dynamically terminated
and the field(s) begins to oscillate around the minimum of its
effective potential $V(\phi)$ as in chaotic inflation
\cite{linde}, or ``waterfalls'' towards the minimum of $V$ as in
hybrid inflation \cite{hybr}. After inflation the homogeneous
inflaton field(s) decays due to its interactions with other fields
or its self-interaction.  If the inflaton decay into other fields
were slow as in perturbation theory the created particles would
settle into thermal equilibrium as the inflaton decayed. However,
the decay of the inflaton typically occurs via rapid,
non-perturbative mechanisms collectively known as preheating
\cite{KLSearly}.  The character of preheating may vary from model
to model, e.g. parametric excitation in chaotic inflation
\cite{KLS97} and another, specific type of preheating  in hybrid
inflation \cite{hh}, but its distinct feature remains the same:
rapid amplification of one or more bosonic fields to exponentially
large occupation numbers. This amplification is eventually shut
down by backreaction of the produced fluctuations. The end result
of the process is a turbulent medium of coupled, inhomogeneous,
classical waves far from equilibrium \cite{KT}.

Despite the development of our understanding of preheating after
inflation, the transition from this stage to a  hot Friedmann
universe in thermal equilibrium has remained relatively poorly
understood. A theory of the thermalization of the fields generated
from preheating is necessary to bridge the gap between inflation
and the Hot Big Bang.  The details of this thermalization stage
depend on the constituents of the fundamental Lagrangian ${\cal
L}(\phi_i, \chi_i, \psi_i, A_{\mu}, h_{\mu\nu}, ...)$ and their
couplings, so at first glance it would seem that a description of
this process would have to be strongly model-dependent. We have
found, however, that many features of this stage seem to hold
generically across a wide spectrum of models. This fact is
understandable because the conditions at the end of preheating are
generally not qualitatively sensitive to the details of inflation.
Indeed, at the end of preheating and beginning of the turbulent
stage (denoted by $t_*$), the fields are out of equilibrium. We
have examined many models and found that at $t_*$ there is not
much trace of the linear stage of preheating and conditions at
$t_*$ are not qualitatively sensitive to the details of inflation.
We therefore expect that this second, highly nonlinear, turbulent
stage of preheating may exhibit some universal, model-independent
features.

Although a realistic model would include one or more Higgs-Yang-Mills
sectors, we treat the simpler case of interacting scalars.  Within
this context, however, we consider a number of different models
including several chaotic and hybrid inflation scenarios with a
variety of couplings between the inflaton and other matter fields.

There are many questions about the thermalization process that we
set out to answer in our work. Could the turbulent waves that
arise after preheating be described by the theory of (transient)
Kolmogorov turbulence or would they directly approach thermal
equilibrium? Could the relaxation time towards equilibrium be
described by the naive estimate $\tau \sim (n \sigma_{int})^{-1}$,
where $n$ is a density of scalar particles and $\sigma_{int}$ is a
cross-section of their interaction? If the inflaton $\phi$ were
decaying into a field $\chi$, what effect would the presence of a
decay channel $\sigma$ for the $\chi$ field have on the
thermalization process? For that matter, would the presence of
$\sigma$ significantly alter the preheating of $\chi$ itself, or
even destroy it as suggested in \cite{PR}? How strongly
model-dependent is the process of thermalization; are there any
universal features across different models? Finally there's the
question of chaos. It is known that Higgs-Yang-Mills systems
display chaotic dynamics during thermalization \cite{Biro}. The
possibility of chaos in the case of a single, self-interacting
inflaton was mentioned in passing in \cite{KT}, but when we began
our work it was unclear at what stage of preheating chaos might
appear, and in what way.

Because the systems we are studying involve strong, nonlinear
interactions far from thermal equilibrium it is not possible to solve
the equations of motion using linear analysis in Fourier
space. Instead we solve the scalar field equations of motion directly
in position space using lattice simulations. These simulations
automatically take into account all nonlinear effects of scattering
and backreaction.  Using these numerical results we have been able to
formulate a set of empirical rules that seem to govern thermalization
after inflation.  These rules qualitatively describe thermalization in
a wide variety of models. The features of this process are in some
cases very different from our initial expectations.

Section \ref{setting} gives a brief review of preheating in
different inflationary models. This review should serve to
motivate our study and place it in the broader context of
inflationary cosmology. Sections \ref{results} and \ref{others}
describe the results of our numerical calculations. Section
\ref{results} describes one simple chaotic inflation model that we
chose to focus on as a clear illustration of our results, while
section \ref{others} discusses how the thermalization process
occurs in a variety of other models. Section \ref{chaos} describes
the onset of chaos during preheating and includes a discussion of
the measurement and interpretation of the Lyapunov exponent in
this context. Section \ref{rules} contains a list of empirical
rules that we have formulated to describe thermalization after
preheating. Section \ref{discussion} discusses these results and
other aspects of non-equilibrium scalar field dynamics. Finally,
there is an appendix that describes our lattice simulations.

\section{Inflation and Preheating}\label{setting}

In this section we outline the context where the problem of
thermalization after inflation arises. In the inflationary
scenario, the very early universe expands (quasi)exponentially due
to a vacuum-like equation of state. Such an equation of state can
arise in a number of different  ways, most of which are based on a
homogeneous condensate of one or more classical scalar fields. We
will consider two types of inflationary models. The first is
chaotic inflation \cite{linde} with the single scalar field
potential $V(\phi)$. The second is hybrid inflation, which
involves several scalar fields \cite{hybr}. The properties of
these models are widely discussed in the literature. We will be
dealing only with the decay of the homogenous inflaton condensate
into inhomogeneous modes of the same or other scalars and the
subsequent interactions of these inhomogeneous modes as they
approach thermal equilibrium. Any particles present before or
during inflation are diluted by the exponential expansion. Thus by
the end of inflation all energy is contained in the potential
$V(\phi, ...)$ of one or more classical, slowly moving,
homogeneous inflaton fields. Immediately after inflation the
background field(s) is moving fast and produces particles of the
fields coupled to it. These created particles are mutually
interacting and ultimately must end up in thermal equilibrium.
However, particles may be created so fast that they spend some
time in non-equilibrium states with very large occupation numbers.

Consider chaotic inflation with the potential
\begin{equation} \label{cha_eqn}
V(\phi) = {m^2 \over 2}\phi^2 +   {\lambda \over 4} \phi^4  \, .
\end{equation}
Soon after the end of inflation the homogeneous inflaton field
$\phi(t)$ coherently oscillates around the minimum of its potential
with an amplitude on the order of a Planck mass. The inflaton
oscillations decay due to the creation of particles interacting with
$\phi$. Let $\chi$ be another scalar field coupling with the inflaton
field as ${1 \over 2} g^2 \phi^2 \chi^2$. Particles of the $\chi$
field are produced from the interaction of the quantum vacuum state of
$\chi$ with the coherently oscillating classical field $\phi$. The
dominant channel for this production is the non-perturbative mechanism
of parametric excitation. The $\chi_k$ mode functions exponentially
increase with time as $ \chi_k \simeq {e^{\mu_k t}}$, where the
characteristic exponent $\mu_k$ is a model-dependent function
\cite{KLS97,GKLS}. The copious production of $\chi$ particles
constitutes the first stage of preheating after inflation
\cite{KLSearly}.  This state can be studied with analytical methods
developed in \cite{KLS97,GKLS,B}.  However, very soon the amplitudes
of the inhomogeneous modes (i.e. the occupation number $n_k$) of
$\chi$ become so large that the back-reaction of created particles
must be taken into account. The most important back-reaction effect
will be the rescattering of particles $\chi\phi \to \chi\phi$
\cite{KT}, which is difficult to describe analytically
\cite{KLS97}. Thus, to follow the evolution of the interacting scalar
fields after the first stage of preheating (dominated by parametric
resonance), one must investigate the full non-linear dynamics of the
interacting scalars.

The  Hartree approximation, which is often used for problems of
nonequilibrium quantum field theory, is insufficient here for
several reasons. It fails when field fluctuations have amplitudes
comparable with that of the background field, which occurs
exponentially fast in our case. It does not take into account the
rescattering of particles. Moreover, in the context of preheating
there are diagrams beyond the Hartree approximation that survive
in the $N \to \infty$ limit and give comparable contributions to
those included in the Hartree approximation \cite{KLS97,GKLS}.

Fortunately, scalar fields with high occupation numbers can be
interpreted as classical waves, and the problem can be treated
with lattice simulations  \cite{FT}.
 Such simulations provide
approximate solutions to nonequilibrium quantum  field theory
problems, and we believe they include the leading physical
effects.

Hybrid inflation models involve multiple scalar fields. The simplest
potential for two-field hybrid inflation is
\begin{equation} \label{hyb_eqn}
V(\phi,\sigma) = {\lambda \over 4} (\sigma^2 - v^2)^2 + {g^2 \over
2} \phi^2 \sigma^2  \, .
\end{equation}
Inflation in this model occurs while the homogeneous $\phi$ field
slow rolls from large $\phi$ towards the bifurcation point at
$\phi = {\sqrt{\lambda} \over g} v $ (due to the slight lift of
the potential in $\phi$ direction).  Once $\phi(t)$ crosses the
bifurcation point, the curvature of the $\sigma$ field,
$m^2_{\sigma} \equiv \partial^2 V/\partial \sigma^2$, becomes
negative. This negative curvature results in exponential growth of
$\sigma$ fluctuations. Inflation then ends abruptly in a
``waterfall'' manner. It was recently found \cite{hh}
 that there is strong preheating in
hybrid inflation, but its character is quite different from
preheating based on parametric resonance.

One reason to be interested in hybrid inflation is that it can be
easily implemented in supersymmetric theories. In particular, for
illustration we will use supersymmetric F-term inflation as an example
of a hybrid model.

\section{Calculations in Chaotic Inflation}\label{results}

In this section we present the results of our numerical lattice
simulations of the dynamics of  interacting scalars after
inflation. We discuss in detail one simple model that we have
chosen to illustrate the general properties of thermalization
after preheating. The next section will discuss thermalization in
the context of other models.

\subsection{Model}

The example we have chosen to focus on is chaotic inflation with a
quartic inflaton potential. The inflaton $\phi$ has a four-legs
coupling to another scalar field $\chi$, which in turn can couple
to one or more other scalars $\sigma_i$. The potential for this
model is
\begin{equation}\label{nfldlambda}
V = {1 \over 4} \lambda \phi^4 + {1 \over 2} g^2 \phi^2 \chi^2 +
{1 \over 2} h_i^2 \chi^2 \sigma_i^2.
\end{equation}
The equations of motion for the model (\ref{nfldlambda}) are given by
\begin{equation}\label{phi}
\ddot{\phi} + 3 {\dot{a} \over a} \dot{\phi} -{1 \over a^2}
\nabla^2 \phi + \left(\lambda \phi^2 + g^2 \chi^2\right) \phi = 0
\end{equation}
\begin{equation}\label{chi}
\ddot{\chi} + 3 {\dot{a} \over a} \dot{\chi} -{1 \over a^2}
\nabla^2 \chi + \left(g^2 \phi^2 + h_i^2 \sigma_i^2\right) \chi =
0
\end{equation}
\begin{equation}\label{sigma}
\ddot{\sigma_i} + 3 {\dot{a} \over a} \dot{\sigma_i} -{1 \over
a^2} \nabla^2 \sigma_i + \left(h_i^2 \chi^2\right) \sigma_i = 0.
\end{equation}
We also included self-consistently the evolution of the scale
factor $a(t)$.  The model described by these equations is a
conformal theory, meaning that the expansion of the universe can
be (almost) eliminated from the equations of motion by an
appropriate choice of variables \cite{GKLS}.  See the appendix
for more information on the lattice simulations we used to solve
these equations, including information on the initial conditions
and the rescaled units we used in the calculations and in the
plots we show here.

Preheating in this theory in the absence of the $\sigma_i$ fields
was described in \cite{GKLS}. For $g^2 \gtrsim \lambda$ the field
$\chi$ will experience parametric amplification, rapidly rising to
exponentially large occupation numbers. In the absence of the
$\chi$ field (or for sufficiently small $g$) $\phi$ will be
resonantly amplified through its own self-interaction, but this
self-amplification is much less efficient than the two-field
interaction. The results shown here are for $\lambda = 9 \times
10^{-14}$ (for COBE normalization) and $g^2 = 200 \lambda$. When
we add a third field we use $h_1^2 = 100 g^2$ and when we add a
fourth field we use $h_2^2 = 200 g^2$.

\subsection{The Output Variables}

There are a number of ways to illustrate the behavior of scalar
fields, and different ones are useful for exploring different
phenomena. The raw data is the value of the field $f(t,\vec x)$,
or equivalently its Fourier transform $f_k(t)$. One of the
simplest quantities one can extract from these values is the
variance
\begin{equation}\label{var}
\langle \left(f(t) - \bar f(t)\right)^2 \rangle = {1 \over
{(2\pi)^3}} \int d^3k \vert f_k(t) \vert^2 \, ,
\end{equation}
where the integral does not include the contribution of a possible
delta function at $\vec k = 0$, representing the mean value $\bar
f$.

One of the most interesting variable to calculate is the
(comoving) number density of particles of the $f$-field
\begin{equation}\label{num}
n_f(t) \equiv {1 \over (2\pi)^3} \int d^3k n_k(t)  \, ,
\end{equation}
where $n_k$ is the (comoving) occupation number  of particles
\begin{equation}\label{ad}
n_k(t) \equiv {1 \over 2 \omega_k} \vert \dot f_k\vert^2 + {\omega_k
\over 2} \vert f_k\vert^2
\end{equation}
\begin{equation}
\omega_k \equiv \sqrt{k^2 + m^2_{eff}}
\end{equation}
\begin{equation}\label{mass}
m_{eff}^2 \equiv {\partial^2 V \over \partial f^2} \, .
\end{equation}
For the model (\ref{nfldlambda}) this effective mass is given by
\begin{equation}
m_{eff}^2 = \left\{\begin{array}{ll} 3 \lambda \langle  \phi^2
\rangle + g^2 \langle\chi^2 \rangle
\\g^2 \langle \phi^2 \rangle+ h_i^2  \langle \sigma_i^2 \rangle
\\h_i^2 \langle \chi^2 \rangle \end{array}\right.
\end{equation}
for $\phi$, $\chi$, and $\sigma_i$ respectively. For the classical
waves of $f$ that we are dealing with, $n_k$ corresponds to  an
adiabatic invariant of the waves.
 Formula ($\ref{ad}$) can be interpreted as a
particle occupation number in the limit of large amplitude of the
$f$-field. As we will see below this occupation number spectrum
contains important information about thermalization. Notice that
the effective mass of the particles depends on the variances of
the fields and may be significant and time-dependent. The momenta
of the particles do not necessarily always exceed their masses,
meaning the interacting scalar waves are not necessarily always in
the kinetic regime. In particular this means that in general we
cannot calculate the energies of the fields simply  as $\int
d^3k\, \omega_k n_k$ because interaction terms between fields can
be significant.

From here on we will use $n$ without a subscript to denote the
total number density for a field, and will use the subscript only
to specify a particular field, e.g. $n_\phi$. We use $n_{tot}$ to
mean the sum of the total number density for all fields combined.
Occupation number will always be written $n_k$ and it should be
clear from context which field is being referred to.

In practice it is not very important whether you consider the spectrum
$f_k$ and the variance of $f$ or the spectrum $n_k$ and the number
density. Both sets of quantities qualitatively show the same behavior
in the systems we are considering. The variance and number density
grow exponentially during preheating and evolve much more slowly
during the subsequent stage of turbulence. Most of our results are
shown in terms of number density $n_f$ and occupation number $n_k$
because these quantities have obvious physical interpretations, at
least in certain limiting cases. We shall occasionally show plots of
variance for comparison purposes.

We will follow the evolution of $n(t)$ and $n_k(t)$. The evolution
of the total number density $n_{tot}$ is an indication of the
physical processes taking place. In the weak interaction limit the
scattering of classical waves via the interaction term ${1 \over
2} g^2 \phi^2 \chi^2$ can be treated using a perturbation
expansion with respect to $g^2$. The leading four-legs diagrams
for this interaction corresponds to a two-particle collision
$(\phi \chi \rightarrow \phi \chi)$, which conserves $n_{tot}$.
The regime where such interactions dominate corresponds to ``weak
turbulence'' in the terminology of the theory of wave turbulence
\cite{waves}. If we see $n_{tot}$ conserved it will be an
indication that these two-particle collisions constitute the
dominant interaction. Conversely, violation of $n_{tot}(t)=const$
will indicate the presence of strong turbulence, i.e. the
importance of many-particle collisions. Such higher order
interactions may be significant despite the smallness of the
coupling parameter $g^2$ (and others) because of the large
occupation numbers $n_k$. Later, when these occupation numbers are
reduced by rescattering, the two-particle collision should become
dominant and $n_{tot}$ should be conserved.

For a bosonic field in thermal equilibrium with a temperature $T$
and a chemical potential $\mu$ the spectrum of occupation numbers
is given by
\begin{equation}\label{bose}
n_k = {1 \over {e^{{\omega_k -\mu}\over  T}-1}} \, .
\end{equation}
(We use units in which $\hbar=1$.) Preheating generates large
occupation numbers for which equation (\ref{bose}) reduces to its
classical limit
\begin{equation}\label{wave}
n_k \approx {T \over {\omega_k -\mu}} \, ,
\end{equation}
which in turn reduces to $n_k \propto 1/k$ for $k \gg m,\mu$ and
$n_k \approx const.$ for $k \ll m,\mu$. We will compare the
spectrum $n_k$ to this form to judge how the fields are
thermalizing. Here we consider the chemical potential of an
interacting scalar fields
 as a free parameter.

Unless otherwise indicated all of our results are shown in comoving
coordinates that, in the absence of interactions, would remain
constant as the universe expanded. Note also that for most of our
discussion we consider field spectra only as a function of $\vert \vec
k\vert$, defined by averaging over spherical shells in $k$ space. For
a Gaussian field these spectra contain all the information about the
field, and even for a non-Gaussian field most useful information is in
these averages. This issue is discussed in more detail in section
\ref{chaos}.

\subsection{Results}

\begin{figure}
\leavevmode\epsfxsize=.7\columnwidth \epsfbox{\figdir/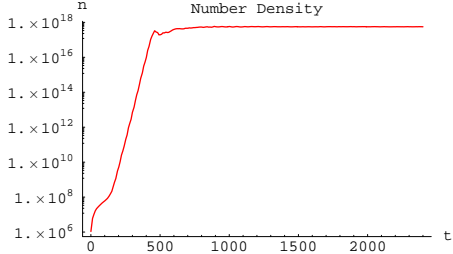}
\caption{$V = 1/4 \lambda \phi^4$. (Note that the vertical scale
is larger than for the subsequent plots.)} \label{n1fldl}
\end{figure}

\begin{figure}
\leavevmode\epsfxsize=.7\columnwidth \epsfbox{\figdir/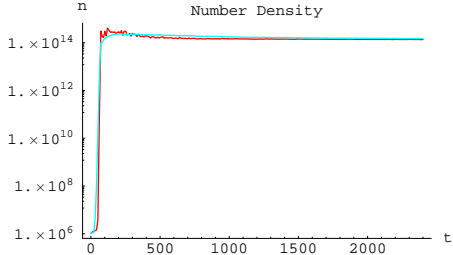}
\caption{$V = 1/4 \lambda \phi^4 + 1/2 g^2 \phi^2 \chi^2$,
$g^2/\lambda=200$. The upper curve represents $n_\chi$.}
\label{n2fldl}
\end{figure}

\begin{figure}
\leavevmode\epsfxsize=.7\columnwidth \epsfbox{\figdir/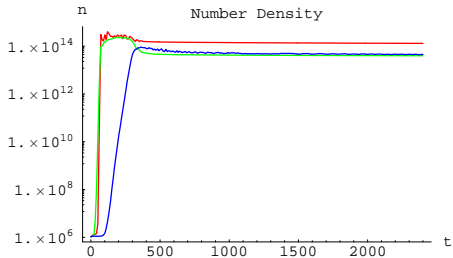}
\caption{$V = 1/4 \lambda \phi^4 + 1/2 g^2 \phi^2 \chi^2 + 1/2 h^2
\chi^2 \sigma^2$, $g^2/\lambda=200$, $h^2 = 100 g^2$. The highest
curve is $n_{\phi}$. The number density of $\chi$ diminishes when
$n_{\sigma}$ grows.} \label{n3fldl}
\end{figure}

\begin{figure}
\leavevmode\epsfxsize=.7\columnwidth \epsfbox{\figdir/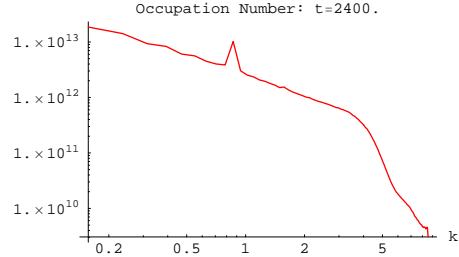}
\caption{$V = 1/4 \lambda \phi^4$.} \label{nk1fldl}
\end{figure}

\begin{figure}
\leavevmode\epsfxsize=.7\columnwidth \epsfbox{\figdir/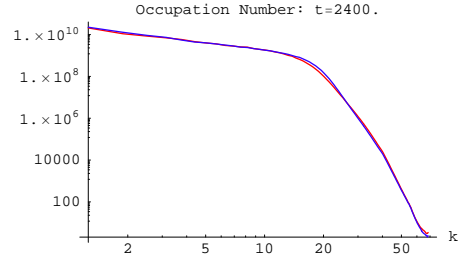}
\caption{$V = 1/4 \lambda \phi^4 + 1/2 g^2 \phi^2 \chi^2$,
$g^2/\lambda=200$. The spectra of $\phi$ and $\chi$ are nearly
identical.} \label{nk2fldl}
\end{figure}

\begin{figure}
\leavevmode\epsfxsize=.7\columnwidth \epsfbox{\figdir/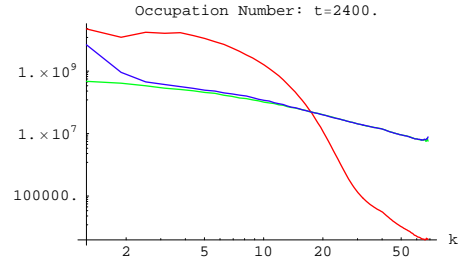}
\caption{$V = 1/4 \lambda \phi^4 + 1/2 g^2 \phi^2 \chi^2 + 1/2 h^2
\chi^2 \sigma^2$, $g^2/\lambda=200$, $h^2 = 100 g^2$ The $\chi$
and $\sigma$ spectra are similar, but $\sigma$ rises in the
infrared). The spectrum of $\phi$ is markedly different from the
others.} \label{nk3fldl}
\end{figure}

\begin{figure}
\leavevmode\epsfxsize=.7\columnwidth \epsfbox{\figdir/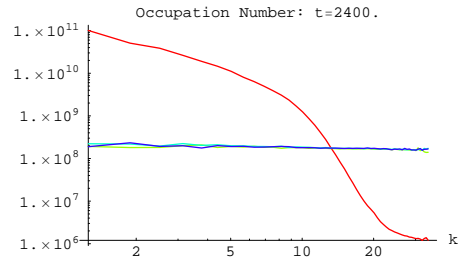}
\caption{$V = 1/4 \lambda \phi^4 + 1/2 g^2 \phi^2 \chi^2 + 1/2
h_i^2 \chi^2 \sigma_i^2$, $g^2/\lambda=200$, $h_1^2 = 200 g^2,
h_2^2 = 100 g^2$. All fields other than the inflaton have nearly
identical spectra.} \label{nk4fldl}
\end{figure}

The key results for this model are shown in
Figures~\ref{n1fldl}-\ref{nk4fldl}, which show the evolution of
$n(t)$ with time for each field and the spectrum $n_k$ for each
field at a time long after the end of preheating. These results
are shown for runs with one field ($\phi$ only), two fields
($\phi$ and $\chi$), and three and four fields (one and two
$\sigma_i$ fields respectively). We will begin by discussing some
general features common to all of these runs, and then comment on
the runs individually.

\begin{figure}
\leavevmode\epsfxsize=.7\columnwidth
\epsfbox{\figdir/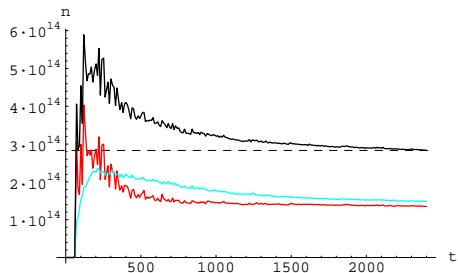} \caption{Number density $n$ for
$V = {1 \over 4} \lambda \phi^4 + {1 \over 2} g^2 \phi^2 \chi^2$.
The plots are, from bottom to top at the right of the figure,
$n_\phi$, $n_\chi$, and $n_{tot}$. The dashed horizontal line is
simply for comparison. The end of exponential growth and the
beginning of turbulence (i.e. the moment $t_*$) occurs around the
time when $n_{tot}$ reaches its maximum.}\label{nlinear2fldl}
\end{figure}

All of the plots of $n(t)$ show an exponential increase during
preheating, followed by a gradual decrease that asymptotically
slows down. See for example Figure~\ref{nlinear2fldl}. This
exponential increase is a consequence of explosive particle
production due to parametric resonance. This regime is fairly well
understood \cite{GKLS}. After preheating the fields enter a
turbulent regime, during which $n(t)$ decreases. This initial,
fast decrease can be interpreted as a consequence of the
many-particle interactions discussed above; as $n_k$ shifts from
low to high momenta the overall number decreases. Realistically,
however, the onset of weak turbulence should be accompanied by the
development of a compensating flow towards infrared modes, which
we would be unable to see because of our finite box size. Thus the
continued, slow decrease in $n(t)$ well into the weak turbulent
regime is presumably a consequence of the lack of very long
wavelength modes in our lattice simulations.

\begin{figure}
\leavevmode\epsfxsize=.7\columnwidth
\epsfbox{\figdir/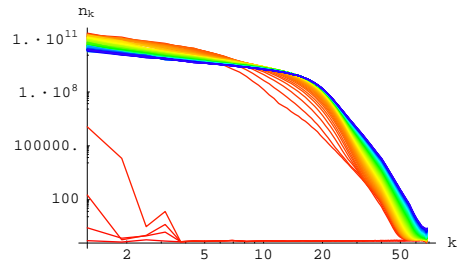} \caption{Evolution of the spectrum
of $\chi$ in the model $V = {1 \over 4} \lambda \phi^4 + {1 \over
2} g^2 \phi^2 \chi^2$. Red plots correspond to earlier times and
blue plots to later ones. For black and white viewing: The sparse,
lower plots all show early times. In the thick bundle of plots
higher up the spectrum is rising on the right and falling on the
left as time progresses.} \label{evolution}
\end{figure}

To see why this shift is occurring look at the spectra $n_k$
(Figures~\ref{nk1fldl}-\ref{nk4fldl}, see also \cite{PR,baryo}).
Even long after preheating Even long after preheating the infrared
portions  of some  of these spectra are tilted more sharply than
would be expected for a thermal distribution (\ref{wave}). Even
more importantly, many of them show a cutoff at some momentum $k$,
above which the occupation number falls off exponentially. Both of
these features, the infrared tilt and the ultraviolet cutoff,
indicate an excess of occupation number at low $k$ relative to a
thermal distribution. This excess occurs because parametric
resonance is typically most efficient at exciting low momentum
modes, and becomes completely inefficient above a certain cutoff
$k_*$. A clear picture of how the flow to higher momenta reduces
these features can be seen in Figure~\ref{evolution}, which shows
the evolution of the spectrum $n_k$ for $\chi$ in the two field
model.

Figure~\ref{evolution} illustrates the initial excitation of modes
in particular resonance bands, followed by a rapid smoothing out
of the spectrum. The ultraviolet cutoff is initially at the
momentum $k_*$ where parametric resonance shuts down, but over
time the cutoff moves to higher $k$ as more modes are brought into
the quasi-equilibrium of the infrared part of the spectrum.
Meanwhile the infrared section is gradually flattening as it
approaches a true thermal distribution. During preheating the
excitation of the infrared modes drives this slope to large,
negative values. From then on it gradually approaches thermal
equilibrium (i.e. a slope of $-1$ to $0$ depending on the chemical
potential  and the mass). The relaxation time for the equilibrium
is significantly shorter than that given by formula $1/n
\sigma_{int}$.  This estimate is valid for dilute gases of
particles, but in our case the large occupation numbers amplify
the scattering amplitudes \cite{KLS97}.

\begin{figure}
\leavevmode\epsfxsize=.7\columnwidth
\epsfbox{\figdir/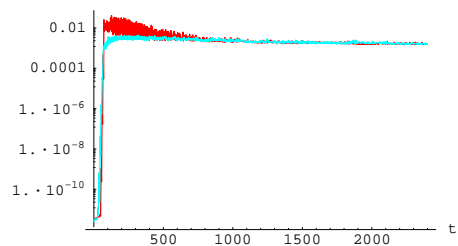} \caption{Variances for $V = {1
\over 4} \lambda \phi^4 + {1 \over 2} g^2 \phi^2 \chi^2$. The
upper plot shows $\langle\left(\phi-\bar{\phi}\right)^2\rangle$
and the lower plot shows
$\langle\left(\chi-\bar{\chi}\right)^2\rangle$.} \label{var2fldl}
\end{figure}

Figure~\ref{var2fldl} shows the evolution of the variances
$\langle\left(f-\bar{f}\right)^2\rangle$ for the two field model.
As indicated above it shows all the same qualitative features as
the evolution of $n$ for that model.

We can now go on to point out some differences between the models,
i.e. between runs with different numbers of fields. The one field
model (pure $\lambda \phi^4$) shows the basic features discussed
above, but the tilt in the spectrum is still very large at the end of
the simulation and $n_\phi$ is decreasing very slowly compared to the
spectral tilt and change in $n$ we see in the two field case. This
difference occurs because the interactions between $\phi$ and $\chi$
greatly speed up the thermalization of both fields. In the one field
case $\phi$ can only thermalize via its relatively weak
self-interaction.

The spectra in the two field run also show a novel feature, namely
that the spectra for $\phi$ and $\chi$ are essentially identical,
 which means among other things
\begin{equation}\label{two}
n_{\phi} \approx  n_{\chi} .
\end{equation}
This matching of the two spectra occurs shortly after preheating
and from then on the two fields evolve identically (except for the
remaining homogeneous component of $\phi$). Aposteriori this
result can be understood as follows. Looking at the potential
$\lambda \phi^4 + g^2 \phi^2 \chi^2$, the second term dominates
because of the hierarchy of coupling strengths $g^2=200\lambda$.
 So the potential $V \approx g^2 \phi^2 \chi^2$ is symmetric
with respect to the two fields, and therefore they act as a single
effective field.

Figures~\ref{n3fldl} and \ref{nk3fldl} show the effects of adding
an additional decay channel for $\chi$.  The interaction of $\chi$
and $\sigma$ does not affect the preheating of $\chi$, but does
drag $\sigma$ exponentially quickly into an excited state.
 The field $\sigma$ is exponentially amplified not by
parametric resonance, but by its stimulated interactions with the
amplified $\chi$ field. Unlike amplification by preheating, this
direct decay nearly conserves particle number, with the result
that $n_\chi$ decreases as $\sigma$ grows, and the spectra of
$\phi$ and $\chi$ are no longer identical. Instead $\chi$ and
$\sigma$ develop nearly identical spectra,
\begin{equation}\label{three}
n_{\chi} \approx  n_{\sigma}< n_{\phi} ,
\end{equation}
and they both thermalize (together) much more rapidly than $\chi$
did in the absence of $\sigma$. There is a looser relationship $
n_{\phi} \approx n_{\sigma}+n_{\chi}$, whose accuracy depends on
the couplings. The inflaton, meanwhile, thermalizes much more
slowly; note the low $k$ of the cutoff in the $\phi$ spectrum in
Figure~\ref{nk3fldl}. By contrast, there is no visible cutoff in
the spectra of $\chi$ and $\sigma$ and the tilt is relatively
mild. The most striking property of this chain of interaction is
the grouping of fields; $\chi$ and $\sigma$ behave identically to
each other and differently from $\phi$. This again can be
understood by the hierarchy of coupling constants,
$h^2=100g^2=20,000\lambda$. The term $h^2 \chi^2 \sigma^2$ is
dominant and puts $\chi$ and $ \sigma$ on an equal footing.

Varying the coupling $h$ did not change the overall behavior of
the system, but it changed the time at which $\sigma$ grew. In the
limiting case $h \gg g$, $\sigma$ grew with $\chi$ during
preheating and remained indistinguishable from it right from the
start. (We found this, for example, for $h^2 = 10,000 g^2$.)

When we added a second $\sigma$ field we found that the $\sigma$
field most strongly coupled to $\chi$ would grow very rapidly and
the more weakly coupled one would then grow relatively slowly.
Note for example that $n_{\sigma 2}$ in Figure~\ref{nk4fldl} grows
more slowly than $n_\sigma$ in Figure~\ref{nk3fldl} despite the
fact that they have the same coupling to $\chi$. In the four field
case $n_\chi$ is reduced when the more strongly coupled $\sigma$
field grows and this slows the growth of the more weakly coupled
one. Nonetheless, the addition of another $\sigma$ field once
again sped up the thermalization of $\chi$ and the $\sigma$
fields. The three fields $\chi$, $\sigma_1$, and $\sigma_2$ once
again have identical spectra
\begin{equation}\label{four}
n_{\chi} \approx  n_{\sigma1}  \approx  n_{\sigma2}  < n_{\phi} ,
\end{equation}
but in the four field case by the end of the run they look
indistinguishable from thermal spectra. If there is an ultraviolet
cutoff for these spectra it is at momenta higher than can be seen
on the lattice we were using. Again, we notice a  loose
relationship $ n_{\phi} \approx n_{\chi}+n_{\sigma1} +n_{\sigma2}
$. in this case.

We close this section with a few words about the effective masses
of the fields, equation~(\ref{mass}). All the masses are scaled in
the comoving frame, i.e.\ we consider $a^2 m_{eff}^2$, and $m$ is
measured in units of momentum (see appendix).
Figure~\ref{mass2fldl} shows the evolution of the effective masses
in the two field model. Note that the vertical axis of these plots
is in the same comoving units as the horizontal ($k$) axes of the
spectra plots. Since the momentum cutoff was of order $k \sim
5-10$ (see Figure~\ref{evolution}) the mass of $\phi$ was
consistently smaller than the typical momenta of the field. By
contrast $m_\chi$ started out much larger and only gradually
decreased. The fluctuations of $\chi$ remained  massive through
preheating (although with a physical mass $\sim 1/a$) and for
quite a while afterwards the typical momentum of these
fluctuations was $k \sim m$.

\begin{figure}
\leavevmode\epsfxsize=.7\columnwidth
\epsfbox{\figdir/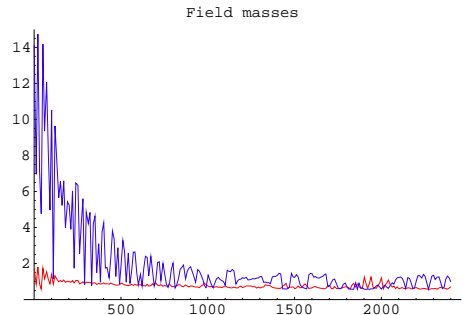} \caption{Effective masses for $V =
{1 \over 4} \lambda \phi^4 + {1 \over 2} g^2 \phi^2 \chi^2$ as a
function of time in units of comoving momentum. The lower plot is
$m_\phi$ and the upper one is $m_\chi$.
 } \label{mass2fldl}
\end{figure}

\begin{figure}
\leavevmode\epsfxsize=.7\columnwidth
\epsfbox{\figdir/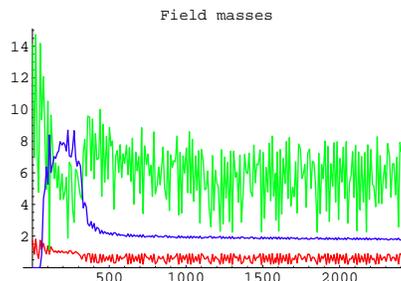} \caption{Time evolution of the
effective masses for the model $V = {1 \over 4} \lambda \phi^4 +
{1 \over 2} g^2 \phi^2 \chi^2 + {1 \over 2} h^2 \chi^2 \sigma^2$.
From bottom to top on the right hand side the plots show $m_\phi$,
$m_\sigma$, and $m_\chi$.} \label{mass3fldl}
\end{figure}

Figure~\ref{mass3fldl} shows the evolution of the effective masses
for the three field model. Once again $m_\phi$ remains small.
Although $m_\sigma$ grows large briefly it quickly subsides.
However, $m_\chi$, with contributions from $\sigma$ and $\phi$,
remains relatively large. Note, however, that the spectrum of
$\chi$ has no clear cutoff after $\sigma$ has grown, so it is
difficult to say whether this mass exceeds a ``typical'' momentum
scale or not.

\section{Other Models of Inflation and Interactions}\label{others}

The model (\ref{nfldlambda}) was chosen to illustrate our basic
results because $\lambda \phi^4$ inflation and preheating is
relatively simple and well studied. Our main interest, however, is
in universal features of thermalization. In this section we
therefore more briefly discuss our results for a variety of other
models. First we continue with $\lambda \phi^4$ inflation by
discussing variants on the interaction potential described above.
Next we discuss thermalization in $m^2 \phi^2$ models of chaotic
inflation. Finally we discuss hybrid inflation.

\subsection{Variations on Chaotic Inflation With a Quartic Potential}

\begin{figure}
\leavevmode\epsfxsize=.7\columnwidth \epsfbox{\figdir/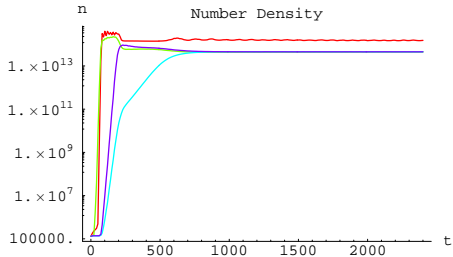}
\caption{$V = 1/4 \lambda \phi^4 + 1/2 g^2 \phi^2 \chi^2 + 1/2
h_i^2 \chi^2 \sigma_i^2$, $g^2/\lambda=200$, $h_1^2 = 200 g^2,
h_2^2 = 100 g^2$. The pattern is similar to the three-field case
until the growth of $\sigma_2$.} \label{n4fldl}
\end{figure}

We looked at several simple variants of the potential
(\ref{nfldlambda}). We considered a model with a further decay
channel for $\sigma$ so that the total potential was
\begin{equation}\label{17}
V = {1 \over 4} \lambda \phi^4 + {1 \over 2} g^2 \phi^2 \chi^2 +
{1 \over 2} h_1^2 \chi^2 \sigma^2 + {1 \over 2} h_2^2 \sigma^2
\gamma^2.
\end{equation}
Setting $h_1 = h_2$ we found that for this four field model
 the evolution of the field
fluctuations, spectra, and number density were qualitatively
similar to those in the four field model (\ref{nfldlambda}).  We
found that at late times
\begin{equation}
n_\chi \approx n_\sigma \approx n_\gamma < n_\phi \ .
\end{equation}
The fields $\chi$, $\sigma$, and $\gamma$ formed a group with
nearly identical spectra and evolution and rapid thermalization,
while $\phi$ remained distinct and thermalized more slowly.
Compare these results to the four field model results in
Figures~\ref{n4fldl} and \ref{nk4fldl}.

We also considered parallel decay channels for $\phi$
\begin{equation}
V = {1 \over 4} \lambda \phi^4 + {1 \over 2} g_1^2 \phi^2 \chi^2 +
{1 \over 2} g_2^2 \phi^2 \gamma^2 + {1 \over 2} h^2 \chi^2
\sigma^2.
\end{equation}
Setting $g_1 = g_2$ and $h^2 = 100 g_1^2$ we found that at late =
times
\begin{equation}
n_\phi \approx n_\gamma > n_\chi \approx n_\sigma \ .
\end{equation}
In other words the four fields formed into two groups of two, with
each group having a characteristic number density evolution.

Finally we looked at adding a self-interaction term for $\chi$
\begin{equation}
V = {1 \over 4} \lambda_\phi \phi^4 + {1 \over 2} g^2 \phi^2
\chi^2 + {1 \over 4} \lambda_\chi \chi^4
\end{equation}
with $\lambda_\chi=g^2$ and found the results were essentially
unchanged from those of the two field runs with no $\chi^4$ term.
The $\chi$ self-coupling caused the spectra of $\phi$ and $\chi$
to deviate slightly from each other, but their overall evolution
proceeded very similarly to the case with no $\chi$
self-interaction term.

\subsection{Chaotic Inflation with a Quadratic Potential}

\begin{figure}
\leavevmode\epsfxsize=.7\columnwidth \epsfbox{\figdir/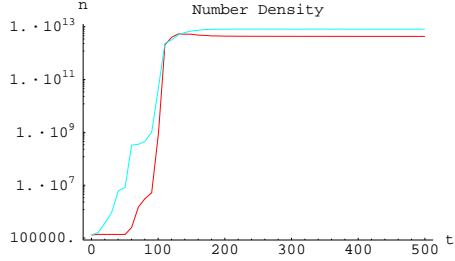}
\caption{$V = 1/2 m^2 \phi^2 + 1/2 g^2 \phi^2 \chi^2$,
$g^2M_p^2/m^2=2.5 \times 10^5$. The upper curve represents
$n_\chi$.} \label{n2fldm}
\end{figure}

\begin{figure}
\leavevmode\epsfxsize=.7\columnwidth \epsfbox{\figdir/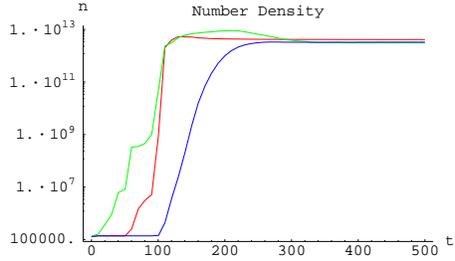}
\caption{$V = 1/2 m^2 \phi^2 + 1/2 g^2 \phi^2 \chi^2 + 1/2 h^2
\chi^2 \sigma^2$, $g^2M_p^2 /m^2=2.5 \times 10^5$, $h^2 = 100
g^2$. The highest curve is $n_{\chi}$. The field that grows latest
is $\sigma$.} \label{n3fldm}
\end{figure}

\begin{figure}
\leavevmode\epsfxsize=.7\columnwidth \epsfbox{\figdir/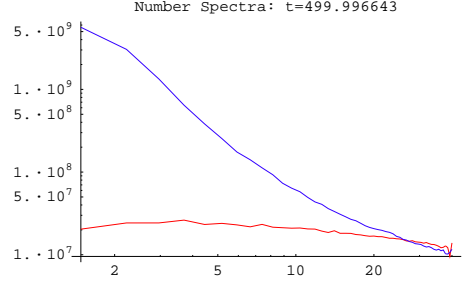}
\caption{$V = 1/2 m^2 \phi^2 + 1/2 g^2 \phi^2 \chi^2$,
$g^2M_p^2/m^2=2.5 \times 10^5$. The upper curve represents the
spectrum of $\chi$.} \label{nk2fldm}
\end{figure}

\begin{figure}
\leavevmode\epsfxsize=.7\columnwidth \epsfbox{\figdir/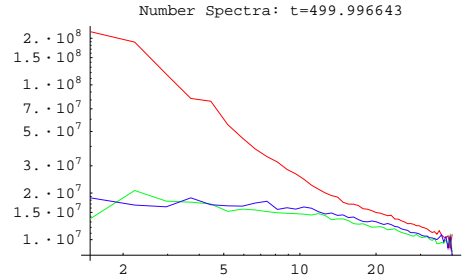}
\caption{$V = 1/2 m^2 \phi^2 + 1/2 g^2 \phi^2 \chi^2 + 1/2 h^2
\chi^2 \sigma^2$, $g^2M_p^2/m^2=2.5 \times 10^5$, $h^2 = 100 g^2$
The $\chi$ and $\sigma$ spectra are similar, while the spectrum of
$\phi$ rises much higher in the infrared.} \label{nk3fldm}
\end{figure}

We also considered chaotic inflation models with an $m^2 \phi^2$
inflaton potential. Figures~\ref{n2fldm}-\ref{nk3fldm} show
results for the model
\begin{equation}\label{nfldm}
V = {1 \over 2} m^2 \phi^2 + {1 \over 2} g^2 \phi^2 \chi^2 + {1
\over 2} h^2 \chi^2 \sigma^2 \, ,
\end{equation}
with $m = 10^{-6} M_p \approx 1.22 \times 10^{13} GeV$ (for COBE),
$g^2 = 2.5 \times 10^5 m^2/M_p^2$, and $h^2 = 100 g^2$.  (See the
appendix for more details.) We considered separately the case of
two fields $\phi$ and $\chi$ and three fields $\phi$, $\chi$, and
$\sigma$. This model exhibits parametric resonance similar to the
resonance in quartic inflation \cite{KLS97}, which results in the
rapid growth of $n$ seen in these figures. The spectra produced in
this way are once again tilted towards the infrared. In the two
field case, $\phi$ and $\chi$ do not have identical spectra as
they did for quartic inflation. This is because the coupling term
$1/2 g^2 \phi^2 \chi^2$ redshifts more rapidly than the mass term
$1/2 m^2 \phi^2$, so the latter remains dominant in the potential,
which is therefore not symmetric between $\phi$ and $\chi$. In the
three field case we again see similar spectra for $\chi$ and
$\sigma$, although they are not as indistinguishable as they were
in $\lambda \phi^4$ theory. The basic features of rapid growth of
$n$, high occupation of infrared modes, and then a flux of number
density towards ultraviolet modes and a slow decrease in $n_{tot}$
are all present as they were for $\lambda \phi^4$ theory. The
shape of the $\phi$ spectrum does not appear thermal, but it is
unclear if this spectrum is compatible with Kolmogorov turbulence.

\subsection{Hybrid Inflation}\label{hybrid}

Preheating has been studied in many different versions of hybrid
inflation, mostly only at the early stages when the equations for
the fluctuations can be linearized. It had been thought until
recently that preheating was not a universal process in hybrid
inflation. In our recent study \cite{hh}, however, we found that
there is generally a very strong preheating in hybrid models,
 but its character is quite
different from preheating based on parametric resonance.  We
discuss in detail in a separate publication \cite{hh} our recent
analytical and numerical studies of preheating in hybrid inflation
models, including a simple two-field model (\ref{hyb_eqn})
 as well as more complex
SUSY F-Term and D-Term models.  As with parametric resonance, the
result of the  instability is the exponential growth of
long-wavelength modes of the fields.

In this paper we are mostly interested in preheating in the
non-inflaton sector and the nonlinear stage after preheating.  In
\cite{hh} we studied the instability  in the inflaton
sector of the hybrid model, i.e. the decay of the homogeneous
fields and excitations of their fluctuations. Here we take a
complementary approach and consider the dynamics of the model with
an additional scalar field $\chi$ coupled to the fields of the
hybrid inflation model. The potential is
\begin{equation} \label{susy}
V = {\lambda\over4} |4\bar\Sigma\Sigma - v^2|^2 + 4\lambda
|\Phi|^2 \left(|\Sigma|^2 + |\bar{\Sigma}|^2\right) + h^2\chi^2
|\Sigma|^2 \, ,
\end{equation}
where $\lambda = 2.5 \times 10^{-5}$ and $h^2 = 2 \lambda$. Here
$\Phi$, $\Sigma$ and $\bar{\Sigma}$ are the complex scalar fields
of the inflaton sector and $\chi$ is an additional matter field.
Inflation occurs along of the $\Phi$ direction for $\langle\Phi
\rangle \gg v$, when $\Sigma=\bar{\Sigma}=0$.  When the magnitude
of the slow-rolling field $\Phi$ reaches the value
$\langle|\Phi_{\rm c}|\rangle = {v\over2}$ spontaneous symmetry
breaking occurs and the $\Sigma$ fields become excited.  It can be
shown that at the end of inflation and the start of symmetry
breaking the complicated potential (\ref{susy}) can be effectively
reduced to the simple two field potential (\ref{hyb_eqn}) (where
$\phi$ and $\sigma$ are combinations of $\Phi$ and $\Sigma$,
$\bar{\Sigma}$ and $g^2={1 \over 2}\lambda $) plus the coupling
term with $h^2\chi^2 |\Sigma|^2$.

\begin{figure}
\leavevmode\epsfxsize=.8\columnwidth
\epsfbox{\figdir/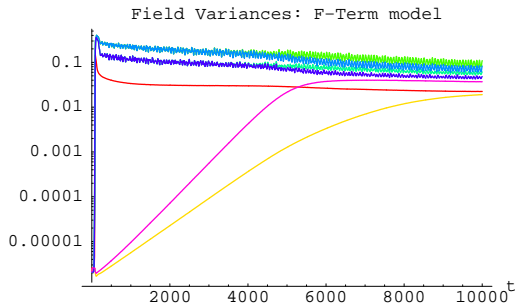} \caption{Evolution of variances of
fields in the model (\ref{susy}). The two fields that grow at late
times, in order of their growth, are $\chi$ and $Im(\Phi)$.}
\label{hybridvars}
\end{figure}

Figure~\ref{hybridvars} show the evolution of the six degrees of
freedom of the inflaton sector as well as the field $\chi$. We see
that all of the inflaton fields except $Im(\Phi)$ are excited very
quickly. Later the fields $\chi$ and $Im(\Phi)$ are dragged into
excited states as well. This dragging corresponds to preheating in
the non-inflaton sector. The fields $\chi$ and $Im(\Phi)$ are
excited by their stimulated interactions with the rest of the
fields. The result of this amplification is a turbulent state that
evolves towards equilibrium very similarly to the chaotic models.
Although the details of inflation and preheating are very
different in hybrid and chaotic models, we found that once a
matter field has been amplified, the thermalization process
proceeds in the same way.

\section{The Onset of Chaos, Lyapunov Exponents and Statistics}\label{chaos}

Interacting waves of scalar fields constitute a dynamical system,
meaning there is no dissipation and the system can be described by
a Hamiltonian. Dynamical chaos is one of the features of wave
turbulence. In this section we address the question if, how and
when the onset of chaos takes place after preheating.

The scalar field fluctuations produced during preheating are
generated in squeezed states \cite{squeezed,KLS97} that are
characterized by correlations of phases between modes ${\vec k}$
and ${-\vec k}$. Because of their large amplitudes we can consider
these fluctuations to be standing classical waves with definite
phases. During the linear stage of preheating, before interactions
between modes becomes significant, the evolution of these waves
may be or may not show chaotic sensitivity to initial conditions.
Indeed, for wide ranges of coupling parameters parametric
resonance has stochastic features \cite{KLS97,GKLS},
 and the issue of the
numerical stability of parametric resonance has not been
investigated. When interaction (rescattering) between waves
becomes important, the waves become decoherent.  At this stage
the waves have well defined occupation numbers but not well
defined phases, and the random phase approximation can be used to
describe the system. This transition signals the onset of
turbulence, following which the system will gradually evolve
towards thermal equilibrium.

 To investigate the
onset of chaos in this system we have to follow the time evolution
of two initially nearby points in the phase space, see e.g.
\cite{Biro}. Consider two configurations of a scalar field $f$ and
$f'$ that are identical except for a small difference of the
fields at a set of points $x_A$. We use $f(t, {\vec x_A}), \dot
f(t, {\vec x_A})$ to indicate the unperturbed field amplitude and
field velocity at the point ${\vec x_A}$ and $f'(t, {\vec x_A}),
\dot f'(t, {\vec x_A})$ to indicate slightly perturbed values at
this point. In other words, the field configurations with $f(t,
{\vec x_A}), \dot f(t, {\vec x_A})$ and $f'(t, {\vec x_A}), \dot
f'(t, {\vec x_A})$ are initially close points in the field phase
space. We then independently evolve these two systems (phase space
points) and observe how the perturbed field values diverge from
the unperturbed ones. Chaos can be defined as the tendency of such
nearby configurations in phase space to diverge exponentially over
time. This divergence is parametrized by the Lyapunov exponent for
the system, defined as
\begin{equation}
\lambda \equiv {1 \over t} log {D(t)\over D_0}
\end{equation}
where $D$ is a distance between two configurations and $D_0$ is
the initial distance at time $0$. Here we define the distance
$D(t)$ simply as
\begin{equation}\label{distance}
D(t)^2 \equiv \sum_A (\vert f_A' - f_A\vert)^2 + (\vert \dot f_A'
- \dot f_A \vert)^2 \ ,
\end{equation}
where we define $f_A \equiv f(t, {\vec x_A})$ and the summation is
taken over all the points where the configurations initially
differed.

\begin{figure}
\leavevmode\epsfxsize=.8\columnwidth
\epsfbox{\figdir/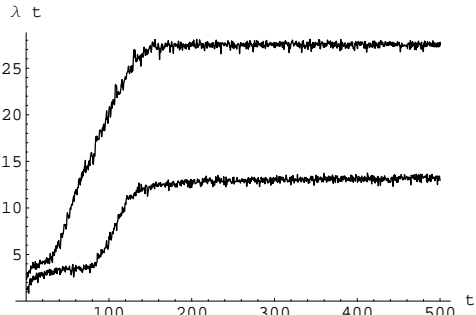} \caption{The Lyapunov exponent
$\lambda$ for the fields $\phi$ (lower curve) and $\chi$ (upper
curve). The vertical axis is $\lambda t$.} \label{lyapunovd}
\end{figure}

For illustration we present the calculations for the model $V = {1
\over 4} \lambda \phi^4 + {1 \over 2} g^2 \phi^2 \chi^2$. We did
two lattice simulations of this model with initial conditions that
were identical except that in one of them we multiplied the
amplitude of $\chi$ by $1 + 10^{-6}$ at 8 evenly spaced points on
the lattice. Figure~\ref{lyapunovd} shows the Lyapunov exponent
for both fields $\phi$ and $\chi$. Note that the vertical axis  is
$\lambda t$ rather than just $\lambda$. During the turbulent stage
the parameter $D(t)$ is artificially saturated to a constant
because of the limited phase space volume of the  system.
Fortunately, the most interesting moment around $t_*$, where the
chaotic motion begins, is covered by this simple approach.
Certainly, the field dynamics continue to be chaotic in subsequent
stages of the turbulence, and one can use more sophisticated
methods to calculate the Lyapunov exponent during these stages
\cite{Lyap,Biro}. However, this issue is less relevant for our
study.

Both fields show roughly the same rate of growth of $\lambda$, but
$\lambda_\chi$ grows much earlier than $\lambda_\phi$ and
therefore reaches a higher level. The reason for this is simple.
The amplitude of $\chi$ is initially very small and grows
exponentially, so even in the absence of chaos we would expect
that during preheating the difference $\chi'(t, {\vec x_A}) -
\chi(t, {\vec x_A})$ must grow exponentially, proportionally to
$\chi \sim e^{\int dt \mu(t)}$ itself. So this exponential growth
is not a true indicator of chaos.

\begin{figure}
\leavevmode\epsfxsize=.8\columnwidth
\epsfbox{\figdir/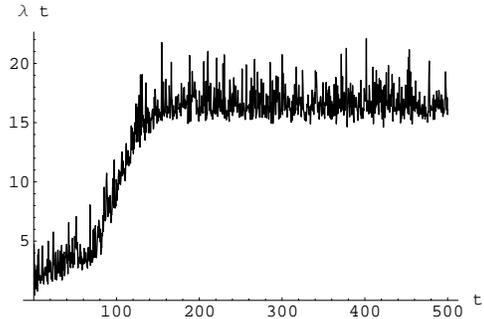} \caption{The Lyapunov exponent
$\lambda'$ for the fields $\phi$ and $\chi$ using the normalized
distance function $\Delta$.} \label{lyapunovdelta}
\end{figure}

To get around this problem and define the onset of chaos in the
context of preheating more meaningfully we introduce a normalized
distance function
\begin{equation}\label{ratio}
\Delta(t) \equiv \sum_A \left({{f'_A-f_A}\over {f'_A+f_A
}}\right)^2 +\left({{\dot f'_A-\dot f_A}\over {\dot f'_A+\dot f_A
}}\right)^2
\end{equation}
that is well regularized even while the field $\chi$ is being
amplified exponentially. Figure~\ref{lyapunovdelta} shows the
Lyapunov exponent $\lambda' \equiv {1 \over t} log {\Delta(t)
\over \Delta(t_0)}$ for $\chi$. In this case we see the onset of
chaos only at the end of preheating. The plot for the $\phi$ field
is nearly identical.  The Lyapunov exponents for the fields were
$\lambda'_\phi \approx \lambda'_\chi  \approx 0.2$ (in the units
of time adopted in the simulation). This corresponds to a very
fast onset of chaos.

Thus we see that chaotic turbulence starts abruptly at the end of
preheating. Initially wave turbulence is strong and rescattering
does not conserve the total number of particles $n_{tot}$. The
fastest variation in $n_{tot}$ occurs at the same time as the
onset of chaos, $t_* \sim 100-200$. We conjecture that the entropy
of the system of interacting waves is generated around the moment
$t_*$. As the particle occupation number drops, the turbulence
will become weak and $n_{tot}$ will be conserved.
Figure~\ref{nlinear2fldl} clearly shows this evolution of the
total number of particles $n_{tot}$ in the model.

\begin{figure}
\leavevmode\epsfxsize=.8\columnwidth
\epsfbox{\figdir/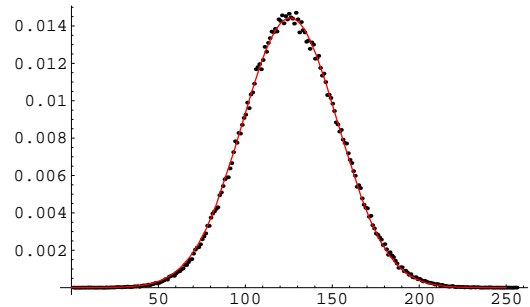} \caption{The probability
distribution function for the field $\chi$ after preheating. Dots
show a histogram of the field and the solid curve shows a best-fit
Gaussian.} \label{gaussian}
\end{figure}

We also considered the statistical properties of the interacting
classical waves in the problem. The initial conditions of our
lattice simulations correspond to random gaussian noise. In
thermal equilibrium, the field velocity $\dot f$ has gaussian
statistics, while the field $f$ itself departs from that unless it
has high occupation numbers. Figure~\ref{gaussian} shows the
probability distribution of the field $\chi$ during the weak
turbulence stage after preheating, and indeed the distribution is
nearly exactly gaussian. Thus, at this stage we can treat the
superposition of classical scalar waves with large occupation
numbers and random phases as random gaussian fields.

\begin{figure}
\leavevmode\epsfxsize=.8\columnwidth
\epsfbox{\figdir/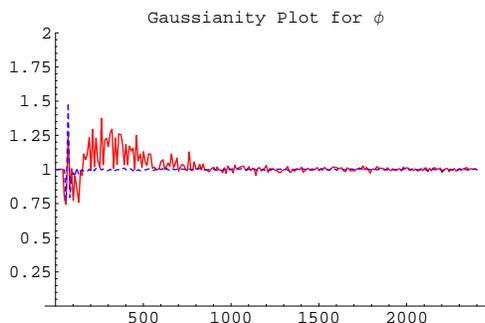} \caption{Deviations from
Gaussianity for the field $\phi$ as a function of time. The solid,
red line shows $3
\langle\delta\phi^2\rangle^2/\langle\delta\phi^4\rangle$ and the
dashed, blue line shows $3
\langle\delta\dot{\phi}^2\rangle^2/\langle\delta\dot{\phi}^4\rangle$.}
\label{phicorrelations}
\end{figure}

\begin{figure}
\leavevmode\epsfxsize=.8\columnwidth
\epsfbox{\figdir/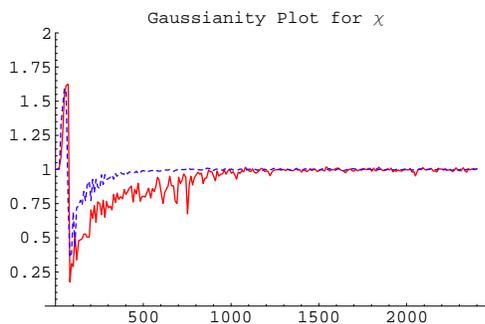} \caption{Deviations from
Gaussianity for the field $\chi$ as a function of time. The solid,
red line shows $3
\langle\delta\chi^2\rangle^2/\langle\delta\chi^4\rangle$ and the
dashed, blue line shows $3
\langle\delta\dot{\chi}^2\rangle^2/\langle\delta\dot{\chi}^4\rangle$.}
\label{chicorrelations}
\end{figure}

During preheating, however, this gaussian distribution is altered.
A simple measure of the gaussianity of a field comes from
examining its moments. For a gaussian field there is a fixed
relationship between the two lowest nonvanishing moments, namely
\begin{equation}
3 \langle\delta\phi^2\rangle^2 = \langle\delta\phi^4\rangle \, ,
\end{equation}
where $\delta\phi \equiv \phi - \langle\phi\rangle$ and angle
brackets denote ensemble averages or, equivalently, large spatial
averages. We measured the ratio of the left and right hand sides
of this equation for $\phi$ and $\chi$ and their time derivatives
using spatial averages over the lattice. The results are shown in
Figures~\ref{phicorrelations} and \ref{chicorrelations}. As
expected, the fields are initially gaussian, deviate from it
during preheating, and rapidly return to it afterwards. The  plots
for the moments of the field velocities are similar, although the
field velocities remain closer to gaussianity.

It is quite important to notice that gaussianity is broken around
the end of preheating and the beginning of the strong turbulence.
In particular, it makes invalid the use of the Hartree
approximation beyond this point.

\section{ Rules of Thermalization}\label{rules}

This paper is primarily an empirical one. We have numerically
investigated the processes of preheating and thermalization in a
variety of models and determined a set of rules that seem to hold
generically. These rules can be formulated as follows:

\bigskip
\noindent {\it 1.  In many, if not all viable models of inflation there
exists a mechanism for exponentially amplifying fluctuations of at
least one field $\chi$. These mechanisms tend to excite
long-wavelength excitations, giving rise to a highly infrared
spectrum.}

The mechanism of parametric resonance in single-field models of
inflation has been studied for a number of years. Contrary to the
claims of some authors, this effect is quite robust. Adding
additional fields (e.g. our $\sigma$ fields) or self-couplings
(e.g. $\chi^4$) has little or no effect on the resonant period.
Moreover, in many hybrid models a similar effect occurs due to
other instabilities. The qualitative features of the fields
arising from these processes seem to be largely independent of the
details of inflation or the mechanisms used to produce the fields.

\bigskip
\noindent {\it 2.  Exciting one field $\chi$ is sufficient to
rapidly drag all other light fields with which $\chi$ interacts
into a similarly excited state.}

We have seen this effect when multiple fields are coupled directly to
$\chi$ and when chains of fields are coupled indirectly to $\chi$. All
it takes is one field being excited to rapidly amplify an entire
sector of interacting fields. These second generation amplified fields
will inherit the basic features of the $\chi$ field, i.e. they will
have spectra with more energy in the infrared than would be expected
for a thermal distribution.

\bigskip
\noindent {\it 3. The excited fields will be grouped into subsets with
identical characteristics (spectra, occupation numbers, effective
temperatures) depending on the coupling strengths.}

We have seen this effect in a variety of models. For example in
the models (\ref{nfldlambda}) and (\ref{17}) the $\chi$ and
$\sigma$ fields formed such a group. In general, fields that are
interacting in a group such as this will thermalize much more
quickly than other fields, presumably because they have more
potential to interact and scatter particles into high momentum
states.

\bigskip
\noindent {\it 4. Once the fields are amplified, they will approach
thermal equilibrium by scattering energy into higher momentum modes.}

This process of thermalization involves a slow redistribution of
the particle occupation number as low momentum particles are
scattered and combined into higher momentum modes. The result of
this scattering is to decrease the tilt of the infrared portion of
the spectrum and increase the ultraviolet cutoff of the spectrum.
Within each field group the evolution proceeds identically for all
fields, but different groups can thermalize at very different
rates.

\section{Discussion}\label{discussion}

We investigated the  dynamics of interacting scalar fields during
post-inflationary preheating and the development of equilibrium
immediately after preheating. We used three dimensional lattice
simulations to solve the non-linear equations of motion of the
classical fields.

There are a number of problems both from the point of view of
realistic models of early universe preheating and from the point
of view of non-equilibrium quantum field theory that we have not
so far addressed. In this section we shall discuss some of them.

Although we considered a series of models of inflation and
interactions, we mostly restricted ourselves to four-legs
interactions. (The sole exception was the hybrid inflation model,
which develops a three-legs interaction after symmetry breaking.)
This meant we still had a residual homogeneous or inhomogeneous
inflaton field. In realistic models of inflation and preheating we
expect the complete decay of the inflaton field. (There are
radical suggestions to use the residuals of the inflaton
oscillations as dark matter or quintessence, but these require a
great deal of fine tuning.) The problem of residual inflaton
oscillations can be easily cured by three-legs interactions. In
the scalar sector three-legs interactions of the type $g^2 v \phi
\chi^2$ may result in stronger preheating. Yukawa couplings $h
\bar \psi \phi \psi$ will lead to parametric excitations of
fermions \cite{GK98}.

There are subtle theoretical issues related to the development of
precise thermal equilibrium in quantum and classical field theory
due to the large number of degrees of freedom, see e.g.
\cite{approx}. In our simulations we see the flattening of the
particle spectra $n_k$ and we describe this as an approach to
thermal equilibrium, but in light of these subtleties we should
clarify that we mean {\it approximate} thermal equilibrium.

Often classical scalar fields in the kinetic regime display
transient Kolmogorov turbulence, with a cascade towards both
infrared and ultraviolet modes \cite{Cond,waves}. In our systems
it appears that the flux towards ultraviolet modes is occurring in
such a way as to bring the fields closer to thermal equilibrium
(\ref{wave}). Indeed, the slope of the spectra $n_k$ at the end of
our simulations is close to $-1$. However, given the size of the
box in these simulations we can little say about the phase space
flux in the direction of infrared modes. This question could be
addressed, for example, with the complementary method of chains of
interacting oscillators, see \cite{Cond}. This is an interesting
problem because an out-of-equilibrium bose-system of interacting
scalars with a conserved number of particles can, in principle,
develop a bose-condensate. It would be interesting to see how the
formation of this condensate would or would not take place in the
context of preheating in an expanding universe. One highly
speculative possibility is that a cosmological bose condensate
could play the role of a late-time cosmological constant.

The highlights of our study for early universe phenomenology are
the following. The mechanism of preheating after inflation is
rather robust and works for  many different systems of
interacting scalars. There is a stage of turbulent classical waves
where the initial conditions for preheating are erased. Initially,
before all the fields have settled into equilibrium with a uniform
temperature, the reheating temperature may be different in
different subgroups of fields. The nature of these groupings is
determined by the coupling strengths.

\section{Acknowledgements}\label{acknowledgements}

The authors are grateful  to Francis Bernardeau, Juan
Garcia-Bellido, Patrick Greene, Andrei Linde and Maxim Lytikov for
useful discussions. This work was supported by NATO Linkage Grant
975389. G.F. thanks CITA for hospitality. The work of G.F. was
also supported by NSF grant PHY-9870115. L.K. was supported by
NSERC and CIAR.

\section*{Appendix: The Lattice Calculations}

All of the numerical calculations reported here were produced with
the program LATTICEEASY, developed by Gary Felder and Igor
Tkachev. The program and documentation are available on the web at
http://physics.stanford.edu/gfelder/latticeeasy/ . The site also
includes all the files needed to implement the particular models
discussed in this paper so anyone can easily reproduce our
results. In this appendix we merely summarize the basics of the
calculation; more details can be found on the website. All
quantities are measured in Planck units ($M_p \approx 1.22 \times
10^{19} GeV$) and we use $f$ to denote a generic scalar field.

The equations of motion for the fields and the scale factor $a$ are
solved on a three-dimensional lattice using finite differencing for
spatial derivatives and a second-order staggered leapfrog algorithm
for time evolution. The evolution equation for a scalar field in an
expanding universe is
\begin{equation}\label{fieldevolution}
\ddot{f} + 3 {\dot{a} \over a} \dot{f} -{1 \over a^2} \nabla^2 f +
{\partial V \over \partial f} = 0
\end{equation}
while the evolution of the scale factor is given by the Friedmann
equations
\begin{equation}\label{friedmann1}
\left({\dot{a} \over a}\right)^2 = {8 \pi \over 3} \rho
\end{equation}
\begin{equation}\label{friedmann2}
\ddot{a} = -{4 \pi \over 3} \left(\rho + 3 p\right) a
\end{equation}
where the energy density and pressure of a scalar field are given by
\begin{equation}
\rho = {1 \over 2} \dot{f}^2 + {1 \over 2} \vert\nabla f\vert^2 +
V
\end{equation}
\begin{equation}
p = {1 \over 2} \dot{f}^2 - {1 \over 6} \vert\nabla f\vert^2 - V.
\end{equation}
In a leapfrog scheme the field values and derivatives are known at
different times, so it is convenient to combine
Equations~(\ref{friedmann1}) and (\ref{friedmann2}) to eliminate
the field derivatives, giving
\begin{equation}
\ddot{a} = -2 {\dot{a}^2 \over a} + {8 \pi \over 3} \left({1 \over
3} \vert \nabla f\vert^2 - a^2 V\right),
\end{equation}
where the gradient is summed over all fields.

The initial conditions were set in momentum space and then Fourier
transformed to give the initial field values on the grid. Starting
at the end of inflation we gave each mode a random phase and a
gaussian distributed amplitude with $rms$ value
\begin{equation}
\langle\vert f_k\vert^2\rangle = {1 \over \sqrt{2 \omega_k}}
\end{equation}
where
\begin{equation}
\omega_k^2 = k^2 + m^2 = k^2 + {\partial^2 V \over \partial f^2}.
\end{equation}

In simulations it's useful to use energy conservation as a check of
accuracy. Energy conservation in an expanding universe is described by
the equation
\begin{equation}\label{rhoevolution}
\dot{\rho} + 3 {\dot{a} \over a} (\rho + p) = 0.
\end{equation}
In principle one could verify that this equation was being
satisfied during the run, but in practice $\dot{\rho}$ is more
difficult to evaluate than $\rho$. Fortunately there is another
way to accomplish the same thing. Equation~(\ref{rhoevolution})
can be derived from the two Friedmann equations (\ref{friedmann1})
and (\ref{friedmann2}), so checking that those two equations are
being simultaneously satisfied is equivalent to checking
Equation~(\ref{rhoevolution}). Since the actual equation for the
evolution of the scale factor is a combination of these two
Friedmann equations we were able to check energy conservation by
calculating the ratio of ${\dot{a} \over a}$ to ${8 \pi \over 3}
\rho$ as the program progressed. (We verified that checking
Equation~(\ref{friedmann2}) gave the same results.) For the
$\lambda \phi^4$ runs the theory is nearly conformal, so almost
the same behavior is obtained with or without the expansion of the
universe (if one uses conformal variables). So we duplicated a
number of our runs without expansion and directly checked energy
conservation. In all cases the results of these two methods of
checking our accuracy were nearly identical. In every run we did,
including cases where we did the run with and without expansion,
energy was conserved to within half a percent over the entire run.

We also did a number of trials to ensure that our results were not
sensitive to our time step, box size, or number of gridpoints.

The field equations were simplified by variable redefinitions. The
redefinitions used and the resulting field equations for the
chaotic inflation models described in the paper are given below.
(Details on the hybrid inflation model can be found in
\cite{hh}.) The units for the fields, times, and momenta in all
the plots in the paper are measured in Planck units rescaled as
indicated below. Before these rescalings, time was in physical
units and distances in comoving coordinates. The momenta $k$ are
also measured in comoving coordinates and they are changed by the
rescalings below as $1/\vec{x}$.

\subsection*{Equations for $\lambda \phi^4$}

For the model (\ref{nfldlambda}) we redefined the field and spacetime
variables as
\begin{equation}
f_{pr} = {a \over \phi_0} f;\;\vec{x}_{pr} = \sqrt{\lambda} \phi_0
\vec{x};\;dt_{pr} = \sqrt{\lambda} \phi_0 {dt \over a}
\end{equation}
where $\phi_0 = .342 M_p$ is the value of the inflaton at the end
of inflation (i.e. at the start of our simulations). This value
was determined from linear numerical calculations as the point at
which ${\partial \phi_{pr} \over \partial t_{pr}}=0$. For $\lambda
= 9 \times 10^{-14}$ one unit of program (conformal) time is $a
(\sqrt{\lambda} \phi_0)^{-1} t_{Planck} \sim a 10^{-36} sec$ and
one unit of program momentum is $a^{-1} \sqrt{\lambda} \phi_0
E_{Planck} \sim a^{-1} 10^{12} GeV$, where $a$ is the scale
factor. In these variables the evolution equations became
\begin{equation}
\phi_{pr}'' - \nabla^2_{pr}\phi_{pr} + \left(\phi_{pr}^2 + {g^2
\over \lambda}\ \chi_{i,pr}^2 - {a'' \over a}\right)\phi_{pr} = 0
\end{equation}
\begin{equation}
\chi_{pr}'' - \nabla^2_{pr}\chi_{pr} + \left({g^2 \over \lambda}
\phi_{pr}^2 + {h_i^2 \over \lambda}\sigma_{i,pr}^2 - {a'' \over
a}\right)\chi_{pr} = 0
\end{equation}
\begin{equation}
\sigma_{i,pr}'' - \nabla^2_{pr}\sigma_{i,pr} + \left({h_i^2 \over
\lambda} \chi_{pr}^2 - {a'' \over a}\right)\sigma_{i,pr} = 0
\end{equation}
\begin{eqnarray}
a'' = -{a'^2 \over a} + {8 \pi \phi_0^2 \over a}\left<{1 \over 3}
\sum_{fields}\left(\vert\nabla_{pr}f_{pr}\vert^2\right) + {1 \over
4} \phi_{pr}^4\right. \nonumber\\ \left.+ {1 \over 2} {g^2 \over
\lambda}\phi_{pr}^2 \chi_{pr}^2 + {1 \over 2} {h_i^2 \over
\lambda}\chi_{pr}^2\sigma_{i,pr}^2\right>
\end{eqnarray}
where primes denote differentiation with respect to $t_{pr}$ and angle
brackets denote spatial averages over the grid.

\subsection*{Equations for $m^2 \phi^2$}

For the model (\ref{nfldm}) we used the following redefinitions
\begin{equation}
f_{pr} = {a^{3/2} \over \phi_0} f;\;\vec{x}_{pr} = m
\vec{x};\;dt_{pr} = m dt
\end{equation}
where in this case $\phi_0 = .193 M_p$.  For $m=10^{-6} M_p$ a
unit of program time corresponded to $m^{-1} T_{Planck} \sim
10^{-30} sec$ and a unit of program momentum corresponded to
$a^{-1} m E_{Planck} \sim a^{-1} 10^{13} GeV$. The evolution
equations became
\begin{eqnarray}
\phi_{pr}'' - a^{-2} \nabla_{pr}^2 \phi_{pr} - {3 \over 4}
\left({a' \over a}\right)^2 \phi_{pr} - {3 \over 2}{a'' \over a}
\phi_{pr} \nonumber\\+ \phi_{pr} + {g^2 \over m^2} \phi_0^2 a^{-3}
\chi_{pr}^2 \phi_{pr} = 0
\end{eqnarray}
\begin{eqnarray}
\chi_{pr}'' - a^{-2} \nabla_{pr}^2 \chi_{pr} - {3 \over 4}
\left({a' \over a}\right)^2 \chi_{pr} - {3 \over 2}{a'' \over a}
\chi_{pr} \nonumber\\+ \phi_0^2 a^{-3} \left({g^2 \over m^2}
\phi_{pr}^2 + {h_i^2 \over m^2} \sigma_{i,pr}^2\right)\chi_{pr} =
0
\end{eqnarray}
\begin{eqnarray}
\sigma_{i,pr}'' - a^{-2} \nabla_{pr}^2 \sigma_{i,pr} - {3 \over 4}
\left({a' \over a}\right)^2 \sigma_{i,pr} - {3 \over 2}{a'' \over
a} \sigma_{i,pr} \nonumber\\+ {g^2 \over m^2} \phi_0^2 a^{-3}
\chi_{pr}^2 \sigma_{i,pr} = 0
\end{eqnarray}
\begin{eqnarray}
a'' = -2{a'^2 \over a} + {8 \pi \phi_0^2\over a^4} \left<{1 \over
3} \sum_{fields}\left(\vert\nabla_{pr}f_{pr}\vert^2\right) + {1
\over 2} a^2 \phi_{pr}^2\right. \nonumber\\ \left.+ {1 \over 2}
\phi_0^2 a^{-1} \left({g^2 \over m^2} \phi_{pr}^2 \chi_{pr}^2 +
{h_i^2 \over m^2} \chi_{pr}^2 \sigma_{i,pr}^2\right)\right>
\end{eqnarray}

\clearpage
\setlength{\columnseprule}{1pt}
\section*{Number Density vs. Time}

\section*{Occupation Number   vs. Momentum}

\clearpage
\setlength{\columnseprule}{1pt}
\section*{Number Density vs. Time}

\newpage
\section*{Number Density vs. Momentum}

\end{document}